\documentclass[conference]{IEEEtran}

\usepackage[nospace]{cite}
\usepackage{xspace}
\usepackage{graphicx}
\usepackage{subfig}
\usepackage{blindtext, xcolor}
\usepackage{comment}
\graphicspath{ {./figures/} }
\usepackage[justification=centering]{caption}
\usepackage[]{algorithmic}
\usepackage{caption}
\usepackage[ruled,vlined]{algorithm2e}
\usepackage{amssymb}
\usepackage{amsmath}

\newcommand{\CASE}[1]{\STATE \textbf{case} #1\textbf{:} \begin{ALC@g}}
\newcommand{\ENDCASE}{\end{ALC@g}}

\newcommand{\DEFAULT}{\STATE \textbf{default:} \begin{ALC@g}}
\newcommand{\ENDDEFAULT}{\end{ALC@g}}
\newcommand{\DEFAULTLINE}[1]{\STATE \textbf{default:} }
\newcommand{\elem}[1]{\textsf{\small{#1}}}
\newcommand{\mRUBiS}{\mbox{mRUBiS}\xspace}
\newcommand{\scalefactor}{0.76}
\usepackage{mathpple}
\usepackage{mathptmx}
\usepackage{mathptm}
\usepackage{newtxmath}
\newcommand{\Pa}[0]{\mathcal{P}}

\usepackage{url}
\makeatletter
\def\ps@IEEEtitlepagestyle{
	\def\@oddfoot{\mycopyrightnotice \thepage}
	\def\@evenfoot{}
}
\def\mycopyrightnotice{
{\footnotesize
	\begin{minipage}{\textwidth}
		\centering
		\textcopyright~2017 IEEE. Personal use of this material is permitted.
		Permission from IEEE must be obtained for all other uses, in any current or future media, including reprinting/republishing this material for advertising or promotional purposes, creating new collective works, for resale or redistribution to servers or lists, or reuse of any copyrighted component of this work in other works. 
		{\tt DOI:} \url{https://doi.org/10.1109/ICAC.2017.35}
	\end{minipage}
}
}
\usepackage[hidelinks]{hyperref}
\hypersetup{
	bookmarks=true,         
	unicode=false,          
	pdftoolbar=true,        
	pdfmenubar=true,        
	pdffitwindow=false,     
	pdftitle={Efficient Utility-Driven Self-Healing Employing Adaptation Rules for Large Dynamic Architectures},
	pdfauthor={Sona Ghahremani, Holger Giese and Thomas Vogel},
	pdfsubject={},
	pdfcreator={},
	pdfproducer={},
	pdfkeywords={self-healing, adaptation rules, architecture-based adaptation, utility},
	pdfnewwindow=true,
}

\begin{document}
\title{Efficient Utility-Driven Self-Healing Employing Adaptation Rules for Large Dynamic Architectures}
\author{\IEEEauthorblockN{Sona Ghahremani, Holger Giese and Thomas Vogel}
\IEEEauthorblockA{Hasso Plattner Institute, University of Potsdam\\
Email: \{sona.ghahremani$|$holger.giese$|$thomas.vogel\}@hpi.uni-potsdam.de}
}

\maketitle
\pagestyle{plain}

\begin{abstract}
Self-adaptation can be realized in various ways. Rule-based approaches prescribe the adaptation to be executed if the system or environment satisfy certain conditions and result in scalable solutions, however, with often only satisfying adaptation decisions. In contrast, utility-driven approaches determine optimal adaptation decisions by using an often costly optimization step, which typically does not scale well for larger problems. 
We propose a rule-based and utility-driven approach that achieves the beneficial properties of each of these directions such that the adaptation decisions are optimal while the computation remains scalable since an expensive optimization step can be avoided. The approach can be used for the architecture-based self-healing of large software systems. We define the utility for large dynamic architectures of such systems based on patterns capturing issues the self-healing must address and we use pattern-based adaptation rules to resolve the issues. 
Defining the utility as well as the adaptation rules pattern-based allows us to compute the impact of each rule application on the overall utility and to realize an incremental and efficient utility-driven self-healing. 
We demonstrate the efficiency and optimality of our scheme in comparative experiments with a static rule-based scheme as a baseline and a utility-driven approach using a constraint solver.
\end{abstract}

\textit{Keywords}: self-healing, adaptation rules, architecture-based adaptation, utility.

\IEEEpeerreviewmaketitle

\section{Introduction}\label{sec:intro}
\noindent
There are various ways how self-adaptation following the MAPE-K feedback loop~\cite{Kephart&Chess2003} and in particular the analyzing and planning phases of the feedback loop can be realized. 

On the one hand, \emph{rule-based} approaches~\cite{1537890,4061119} combine the analyzing and planning phases. Adaptation is executed for specific events and under specific conditions by adaptation rules. In such approaches, events trigger the rules that subsequently check their additional conditions. If the conditions are fulfilled, the actions of the rule can be applied and result in the envisioned changes. The main strengths of rule-based approaches are i) the readability and elegance of individual rules, and ii) the efficiency with which the rules can be processed. The limited expressiveness of the adaptation rules is a drawback. At runtime, the applicable rules are identified (matched) and executed to adapt the system configuration~\cite{1691383}. 

On the other hand, \emph{utility-driven} approaches~\cite{Kephart+Walsh2004,Esfahani+2013} often determine optimal adaptation decisions by using optimization techniques in the planning phase that are guided by a utility function. A utility function determines how valuable each possible system configuration is and the optimization techniques then aim for finding the optimal one. However, the optimization techniques usually prevent that the approaches scale well for large configuration spaces at runtime. To achieve runtime efficiency, linear utility functions are often used since optimizing complex utility functions, as in constraint solver-based approaches, results in non-scalable solutions~\cite{1691383}.

Therefore, we propose in this paper a combined rule-based and utility-driven approach that guarantees optimal adaptation decisions and that is scalable. Thus, the combined approach achieves the individual benefits of both approaches while it avoids the corresponding drawbacks with respect to the optimality of adaptation decisions and scalability.
Particularly, we target the architecture-based self-healing of large software systems and exploit some restrictions usually present for this class of problems to achieve the guarantees for optimality.
We use  our former work to define the utility function in a pattern-based way for large dynamic architectures~\cite{Ghahremani+16} and 
further also define the adaptation rules in a pattern-based way.
This joint use of patterns allows us to combine the utility and the rules and therefore to predict the impact of each rule application on the overall utility.
Based on these predictions for the rules and the knowledge about the costs of applying each adaptation rule, we can determine and execute at runtime the optimal sequence of rule applications. 

We demonstrate these benefits of our approach by comparing it with two alternatives solutions in simulations of \mRUBiS~\cite{mRUBiS}.
We show that our approach is only slightly slower but reaches a higher utility over time (reward) than a static, rule-based solution.
Then, we demonstrate that our approach always makes optimal adaptation decisions similar to an alternative solution using a constraint solver. However, our approach requires considerably less time than the solver, especially for large architectures.
Being incremental makes our approach more scalable since it faces less overhead. As argued by Ghezzi~\cite{Ghezzi2012}, incremental solutions are highly desirable for self-adaptive software. In our earlier work~\cite{Vogel+2009,VogelNHGB10,VG10}, we presented an incremental scheme for the monitoring and execution phases of the feedback loop operating on architectural runtime models. The results of this paper complement the earlier results by enabling the incremental analysis and planning for architectural runtime models based on adaptation rules and utility functions. Therefore, we focus in this paper on the analysis and planning phases of the feedback loop.  

The rest of the paper is structured as follows:
We introduce architectural self-adaptation with runtime models and the pattern-based definition of the utility in Section~\ref{sec:prerequisites}.
Then, we discuss our approach considering its general scheme and its application in a feedback loop in Sections~\ref{sec:scheme} and~\ref{subsec:rulebased}.
We analyze, discuss, and demonstrate the benefits of our approach in Sections~\ref{sec:analysis} and~\ref{sec:evaluation}.
Finally, we discuss related work in Section~\ref{sec:related} and
provide a conclusion with an outlook on future work in Section~\ref{sec:conclusion}.

\section{Prerequisites}\label{sec:prerequisites}

\subsection{Architectural Self-Adaptation and Runtime Models }
\label{subsec:runtime-model}\label{sec:arch-based}

\noindent
To realize self-adaptation, a software system is equipped with a \textit{MAPE-K} feedback loop that \underline{m}onitors and \underline{a}nalyzes the system and if needed, \underline{p}lans and \underline{e}xecutes an adaptation to the system, which is all based on \underline{k}nowledge~\cite{Kephart&Chess2003}.
In this context, many approaches consider the \textit{software architecture} as an appropriate abstraction level (e.g.,~\cite{Oreizy+1999,Garlan+2009,Vogel+2009}) since self-adaptation can be \textit{generally} achieved by adding and removing components as well as connectors among components~\cite{MageeKramer1996}.
For this purpose, the feedback loop maintains a \textit{runtime model}~\cite{Blair+2009}, as part of its knowledge, which represents the architecture of the system under adaptation. The model and the system are \textit{causally connected}, that is, any relevant change of the system is reflected in the model, and vice versa~\cite{Blair+2009}. Thus, the analysis and planning phases can operate on the model.
Technically, a runtime model allows us to employ model-driven engineering (MDE) techniques~\cite{France+Rumpe2007}.
In our earlier work~\cite{VG10,Vogel+2009,VogelNHGB10}, we presented an incremental scheme for the monitor and execute phases that employs MDE techniques and architectural runtime models and that is the basis for this paper.

\begin{figure}[b]
\begin{centering}
\vspace{-7mm}
  \includegraphics[width=0.85\linewidth]{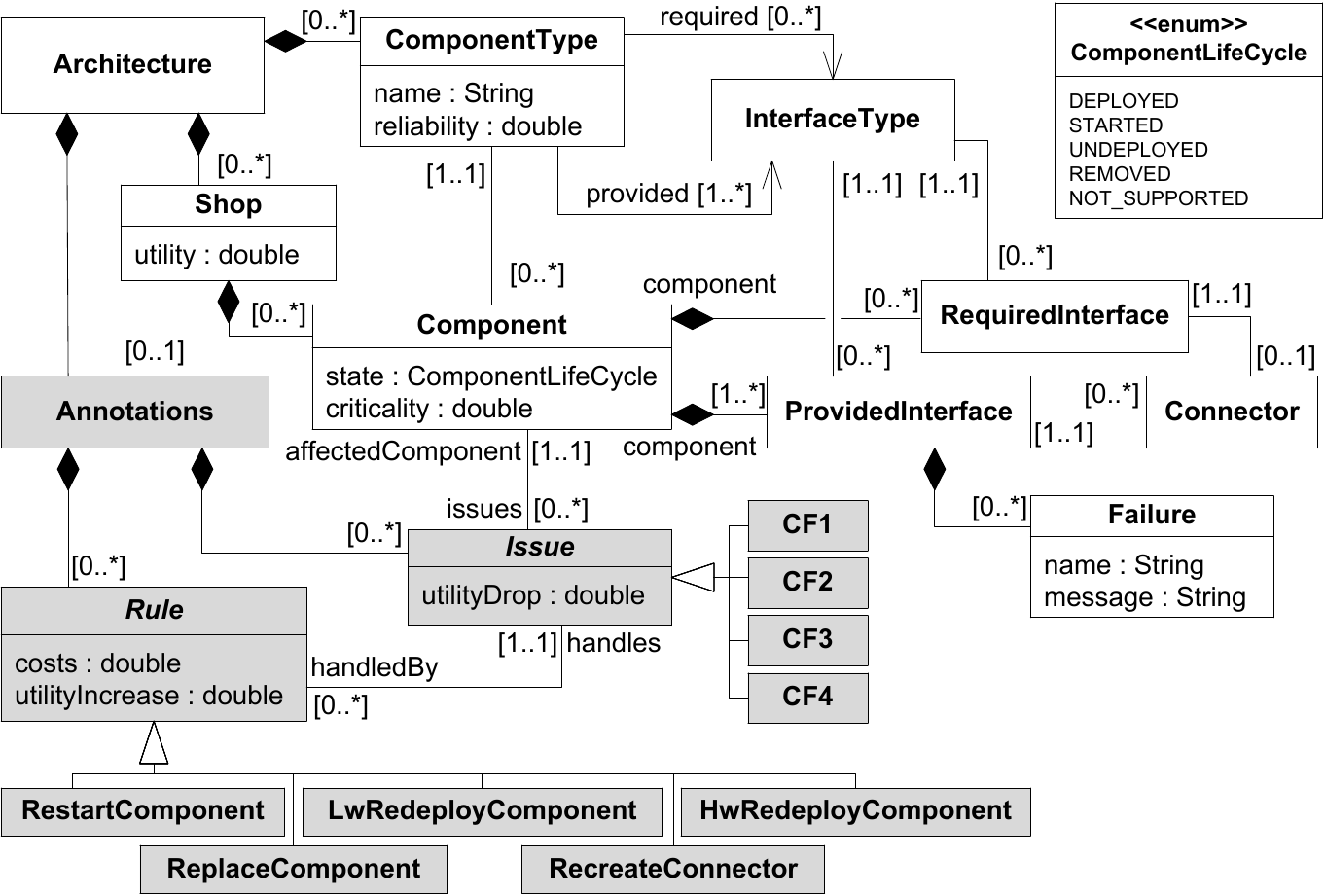}
  \caption{Simplified Metamodel of the Runtime Model.}
  \label{fig:metamodel}
  \end{centering}
\end{figure}

As the running example, we use \textit{\mRUBiS}~\cite{mRUBiS}, a modular variant of RUBiS. \mRUBiS is an online marketplace that hosts an arbitrary number of shops. Each shop consists of 18 components, can be configured differently, and runs isolated from the other shops.
We are particularly interested into self-healing to automatically repair runtime failures by architectural self-adaptation. This allows us to consider \textit{general} repair rules that adapt the architectural configuration of \mRUBiS.
Therefore, we equip \mRUBiS with a \mbox{MAPE-K} feedback~loop that uses an architectural runtime model of \mRUBiS. Specifically, the model represents the runtime architecture of \mRUBiS according to the deployment of \mRUBiS in an EJB application server. This model conforms to the metamodel shown in Fig.~\ref{fig:metamodel}.

The metamodel captures the \mRUBiS \elem{Architecture} with a set of \elem{ComponentType}s that require and provide \elem{InterfaceType}s. For each \elem{Shop}, the same component types are instantiated to \elem{Component}s with their \elem{Provided-} and \elem{RequiredInterface}s. A \elem{Connector} links a required and a provided interface if both are of the same \elem{InterfaceType}. 
These elements allow us to describe the runtime architecture of \mRUBiS. The other elements are relevant for self-adaptation and outlined later.

Using (meta)models and MDE techniques, we realize the analysis rules with model queries and the adaptation rules with in-place model transformations. For self-healing, the analysis rules query the runtime model to identify failures (issues) in \mRUBiS while the adaptation rules determine how to modify the runtime model, that is, how to change the architecture to repair these failures.
To specify a model query, we use a pattern $P$ of a set of patterns $\mathcal{P}$ describing a structural fragment of the architecture $G$. Since the architecture is represented by the runtime model, we also use $G$ to refer to the model. 
An occurrence of a pattern $P$ in the model $G$ corresponds to a match $m$ of $P$ in $G$ (we write $G \models_m P$). For instance, a match identifies a failure in the architecture.
An adaptation rule $r$ in the rule set $\Re$ uses such patterns or already identified matches in the model to localize where adaptation is needed and then to change the model in-place to repair the failure.

\subsection{Pattern-Based Architectural Utility}\label{sec:rbasedutility}\label{subsec:utility}\label{subsec:L-utility}
\noindent
A \emph{utility} function $U$ is an objective policy that expresses how well each configuration of the system in its domain satisfies the functional and non-functional \emph{goals} of the system.  
For this purpose, $U$ assigns a real-value scalar desirability belonging to $[-\infty,+\infty]$ to any possible system configuration $G$. Such scalar values allow us to compare different configurations and to select the one with the highest utility as the best adaptation decision. 
Furthermore, the accumulated utility over time described by the \textit{reward} supports comparisons over time.

Defining a valid utility function is of high importance since in an optimization problem to find the best configuration, it is always the utility function that is maximized not the real utility of the system. 
There has been extensive research on \emph {utility-driven} decision-making policies and elicitation of user preferences (e.g.,~\cite{1317482}). Considering architectural configurations, a typical approach is to predict the impact of each variant of an architectural property on the overall goals. A normalized linear utility function uses these impact values to compute their weighted sum over all properties given a concrete architecture with concrete variants for each property. The weights represent the preferences of the user/developer and the result is the utility of the given architecture~\cite{1128711}. Such an approach can be used for planning a self-adaptation, that is, to identify the target architecture, to which the system should be adapted to.

Moreover, defining utility functions for architectural configurations is challenging, particularly when considering large, dynamic architectures \cite{Cheng2006}.
In the following, we outline our proposal to define utility functions for large, dynamic architectures based on patterns~\cite{Ghahremani+16}. Due to the employed patterns, our utility functions can cope with dynamic architectural changes in contrast to statically defined utility functions. 

We know that for a utility function for architectural runtime models must hold that (i) the optimal architectural configuration where all the system goals are optimally fulfilled must gain the maximum utility and that (ii) if any constraint or goal is violated, this must lead to a decrease of utility. 

According to (i), we include the impact of \emph{present architectural fragments} in the utility.
We define such fragments by \emph{positive architectural utility patterns} $\Pa^+ = \{ P^+_1, \dots, P^+_k \}$ and capture their impact on the utility by utility sub-functions $U_i$.
This impact may vary for each individual occurrence of such a fragment depending on the specific context such as the \elem{criticality} of the concrete components that are present.

\begin{figure}[h]
\vspace{-2mm}
\begin{centering}
 \includegraphics[scale=\scalefactor]{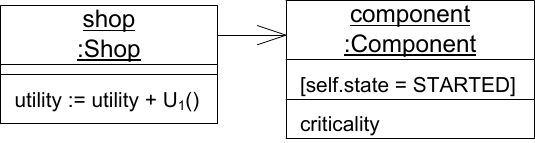}
  \caption{Positive Architectural Utility Pattern $P^+_1$.}
  \label{fig:Pospatt}
  \end{centering}
\vspace{-1mm}
\end{figure}

Fig.~\ref{fig:Pospatt} shows the positive pattern $P^+_1$ and the related utility sub-function $U_1$. The pattern prescribes a started component that is associated to a shop and therefore contributes to the shop's functionality. Matching this pattern for one component in the runtime model, the utility of the associated shop is increased by $U_1$. We define $U_1:=$ criticality of the component $\times$ reliability of the corresponding component type $\times$ connectivity of the component. Matching all components of a shop, the utility of the shop is the sum of the corresponding sub-utilities $U_1$ for all of these components. Finally, the pattern is applied to all shops of \mRUBiS to obtain the utilities of each shop.

Concerning the parts of $U_1$, each component has a \elem{criticality} (see corresponding attribute) denoting its importance for a shop. For instance, the \elem{Authentication} component is more critical than the \elem{Reputation} component since the former is necessarily required by a shop to close a deal while the latter is not.
Additionally, each component type has a \elem{reliability}.
For a certain functionality, alternative component types with different reliabilities exist (e.g., local vs. various third-party authentication services). Thus, selecting the most reliable alternative for a functionality results in a higher utility increase.
Finally, the connectivity of a present component in terms of the number of associated \elem{Connector}s indicates the importance and thus influences the utility increase of the component. 

According to (ii), we further include the negative impact of undesirable situations defined by \emph{negative architectural utility patterns} $\Pa^- = \{ P^-_{k+1}, \dots, P^-_n \}$ in the utility. Such patterns negatively affect the architecture such that they decrease the overall utility according to their utility sub-functions $U_i$. 
Examples of such negative patterns are occurrences of runtime failures.
As before, the impact may vary for each individual occurrence of a negative pattern depending on the specific context such as the \elem{criticality} of the specific components that cause the failures.

\begin{figure}[h]
\vspace{-2mm}
\begin{centering}
  \includegraphics[scale=\scalefactor]{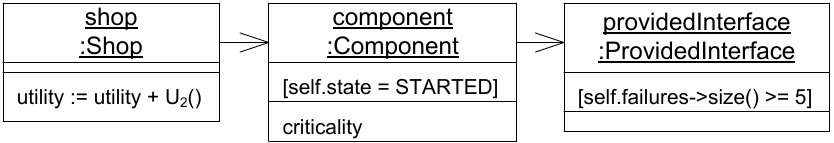}
  \caption{Negative Architectural Utility Pattern $P^-_2$.}
  \label{fig:Antipatt}
  \end{centering}
\vspace{-1mm}
\end{figure}

Fig.~\ref{fig:Antipatt} shows the negative architectural utility pattern $P^-_2$ for \mRUBiS, which prescribes the case when more than four failures (exceptions) occurred in a started component. Each occurrence of such a negative pattern decreases the utility of the associated shop by $U_2$. Note that $U_2$ is typically negative.

Consequently, the positive patterns capture the possible utility gained by the current architectural configuration while the negative patterns capture whether this potential is currently realized. If it is not realized, negative patterns occur in the architecture and correspondingly decrease the utility. In the most extreme case, the whole utility gained by occurrences of positive patterns can be subtracted again when evaluating the negative patterns. In \mRUBiS, adding a new shop always leads to an increase in the utility and is considered a positive pattern while occurrences of failures in components of a shop correspond to a negative pattern and thus reduce the utility. 

\begin{figure}[h]
\begin{centering}
\vspace{-2mm}
 \includegraphics[scale=\scalefactor]{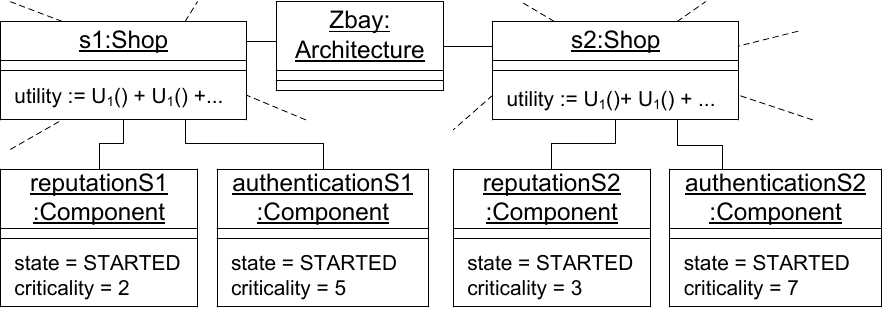}
  \caption{Excerpt of the Architectural Runtime Model.}
  \label{fig:Utility}
  \end{centering}
\vspace{-1mm}
\end{figure}

Fig.~\ref{fig:Utility} shows an excerpt of the runtime model with two matches of the positive pattern $P^+_1$ (see Fig.~\ref{fig:Pospatt}) for each shop, that is, with two started components in each shop. For instance, the elements \elem{s1} and \elem{reputationS1} denote one match and \elem{s1} and \elem{authenticationS1} the other match of $P^+_1$ in shop \elem{s1}.
Each match increases the utility of the shop by $U_1$ taking the characteristics of the specific component into account (e.g., the different criticality values of \elem{reputationS1} and \elem{authenticationS1}). The utility of a shop is the sum of the utility sub-functions $U_1$ for all components of the shop while the utility of \mRUBiS is the sum of the utilities of all shops. 
Similarly, matches for negative patterns would decrease the utility of the shops and therefore of \mRUBiS (not illustrated in Fig.~\ref{fig:Utility}).

Consequently, considering $M_i(G) = \{ m | G \models_m P_i \}$ as the set of matches for the pattern $P_i$ in the current architectural configuration $G$, 
the overall utility function $U(G)$ accumulates all effects due to matches of all patterns $\Pa = \{ P_1, \dots, P_n \}$\footnote{If we do not have to distinguish between positive and negative patterns, we omit the superscript $+$ and $-$ for the patterns $P \in \Pa$.}:
\begin{equation}\label{eq:utility}
  U(G) 
:=
  \sum_{i=1}^{n}
      \sum_{m \in\!M_i\!(G)}
        \!\!\!\!\!U_i(G,m)
\end{equation}
Therefore, adaptation rules can refer to such patterns. On the one hand, they should identify and repair occurrences of negative patterns in the architecture. On the other hand, they should not affect existing but rather enable new occurrences of positive patterns by repairing occurrences of negative patterns.

The definition of the pattern-based utility takes the context into account. Each pattern $P_i$ exactly specifies a context that influences the utility sub-function $U_i$ and thus the increase or decrease of the overall utility. For instance, the pattern $P^+_1$ in Fig.~\ref{fig:Pospatt} specifies the criticality of the component and the associated shop as the context. Similarly, the shop and the component including criticality are the context of the pattern $P^-_2$ in Fig.~\ref{fig:Antipatt}. Moreover, we could extend the context of both patterns, for instance, by taking the component type into account.
When matching a pattern, the concrete context is dynamically identified for each match in the runtime model. Such a concrete context corresponds to a fraction of the runtime model that is navigated to obtain the required information such as the criticality of component to calculate $U_i$ at runtime.

\section{Utility-Driven Rule-Based Adaptation Scheme}\label{sec:scheme}
\noindent
We propose a utility-driven scheme to evaluate dynamic software architectures. As discussed in Section~\ref{subsec:utility}, utility functions are used to map each architectural configuration of a software system to a scalar value indicating how well the configuration satisfies the goals. The need for evaluating \textit{dynamic} architectures is motivated by architectural self-adaptation. If adaptation is required, the feedback loop has to identify a suitable or even the optimal target configuration and select accordingly the adaptation rules that move the system to this configuration. Thus, a feedback loop can use the evaluation scheme to determine the target configuration.
With the proposed scheme, we are particularly interested in self-healing, that is, the automatic repairing of runtime failures by \textit{general} rules that perform architectural adaptation.

In this context, we express \emph{issues} (i.e., runtime failures) for an architecture as model \emph{patterns} such that concrete issues with different {impacts} on the overall utility $U(G)$ relate to occurrences of these patterns in the runtime model $G$. 
Additionally, we can express an adaptation rule $r=(P,\dots$), that is applied on the runtime model if the condition described as a model pattern $P$ is satisfied. We denote for an adaptation rule $r=(P,\dots)$ that an occurrence as a match $m$ for $P$ in the runtime model $G_i$ exists and that applying the rule results in a modified runtime model $G_j$ by $G_i \rightarrow_{r,m} G_j$.

Our scheme can be directly mapped to a MAPE-K feedback loop. 
The \textit{monitoring} phase observes change events emitted by the system to trigger the adaptation. 
During the \textit{analyzing} and \textit{planning}, our scheme requires two decisions: the target configuration of the system, and the rules and their matches that move the system to the target. 
Finally, the last step \textit{executes} these rules for their matches on the running system.

\begin{figure}[t]
	\begin{centering}
		\includegraphics[width=0.45\linewidth]{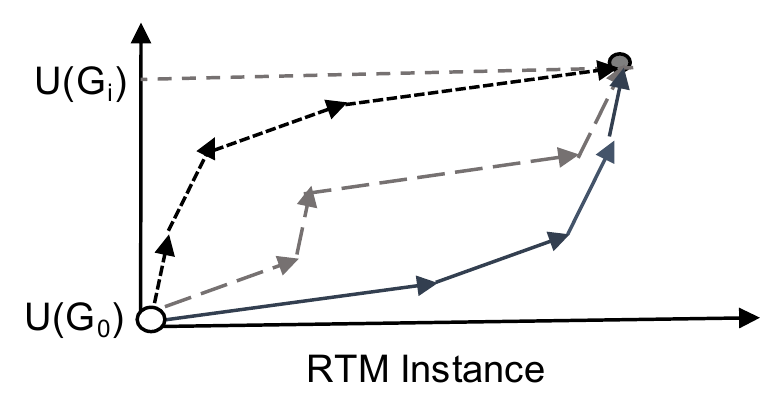}
		\caption{Target Configuration and Different Adaptation Paths.}
		\label{fig:target}
	\end{centering}
\vspace{-4mm}
\end{figure}

These two decisions are inspired by the idea of model-predictive control~\cite{Seborg+2011} that first defines a target and then predicts the optimal path to reach the target. This is illustrated in Fig.~\ref{fig:target} showing one target with three alternative paths to reach the target.
Considering the self-healing, selecting an architecture where issues are repaired is equivalent to defining the target configuration. During the repair, selecting the best sequence of adaptation rules and their matches that resolve all issues is equivalent to building the path toward target.

For the target configuration $G_i$ must hold that the utility $U(G_i)$ must be higher or equal to the utility $U(G_j)$ of all possible next configurations $G_j$ that are the outcomes of resolving the issues in the faulty configuration $G_0$. 
To avoid enumerating the complete search space, our scheme computes the impact of each possible rule application for a match on the related utility sub-function and therefore on the overall utility~($U(G_i)-U(G_0)$).
After defining the target $G_i$, the second step is to select a set of adaptation rules and their matches to actually reach $G_i$. Each rule application changes the runtime model $G$. Starting from a  runtime model instance $G_0$ at the beginning of each \mbox{MAPE-K} cycle, $G_0$ evolves to $G_i$ as a result of a sequence of rule applications: \mbox{$G_0 \rightarrow_{r_1,m_1} G_{1} \rightarrow ...\rightarrow_{r_i,m_i} G_i$}.

Based on the impact of each rule application on the related utility sub-function and thus on the overall utility, we then determine the path. To resolve an issue, multiple rules are applicable and an estimation of their impacts allows us to select a conflict-free subset of them. We assume here that for each set, we can compute $U(G_i)$ regardless of the order in which the rules are executed (see assumption (A4) later). 

Our scheme guarantees
(i) executing the selected set of rules and related matches eventually leads to the target configuration $G_i$ with utility $U(G_i)$ and 
(ii) executing them in the right order results in the highest achievable reward (utility accumulated over time). 
To fulfil~(i), when there are two or more alternative rules and related matches to resolve the same issue, the scheme proceeds with selecting the rule that has the highest impact on the corresponding utility sub-function. 
To achieve~(ii), within a selected conflict-free set of rules, rules are executed in a decreasing order regarding their impact on the corresponding utility sub-functions. 
We claim and show that the proposed approach is optimal regarding the final utility $U(G_i)$ and the achieved reward in the meantime.

\section{Linking Utility to Adaptation Rules}\label{subsec:rulebased}
\noindent
The utility functions for architectural runtime models as we defined them in Section~\ref{subsec:utility} in principle allow us to follow an optimization-based approach that searches the configuration space and computes the utility for each possible configuration. However, such a solution is rather wasteful if the utility would have to be computed for each configuration completely anew.

\begin{figure}[t]
\begin{centering}
  \includegraphics[width=0.8\linewidth]{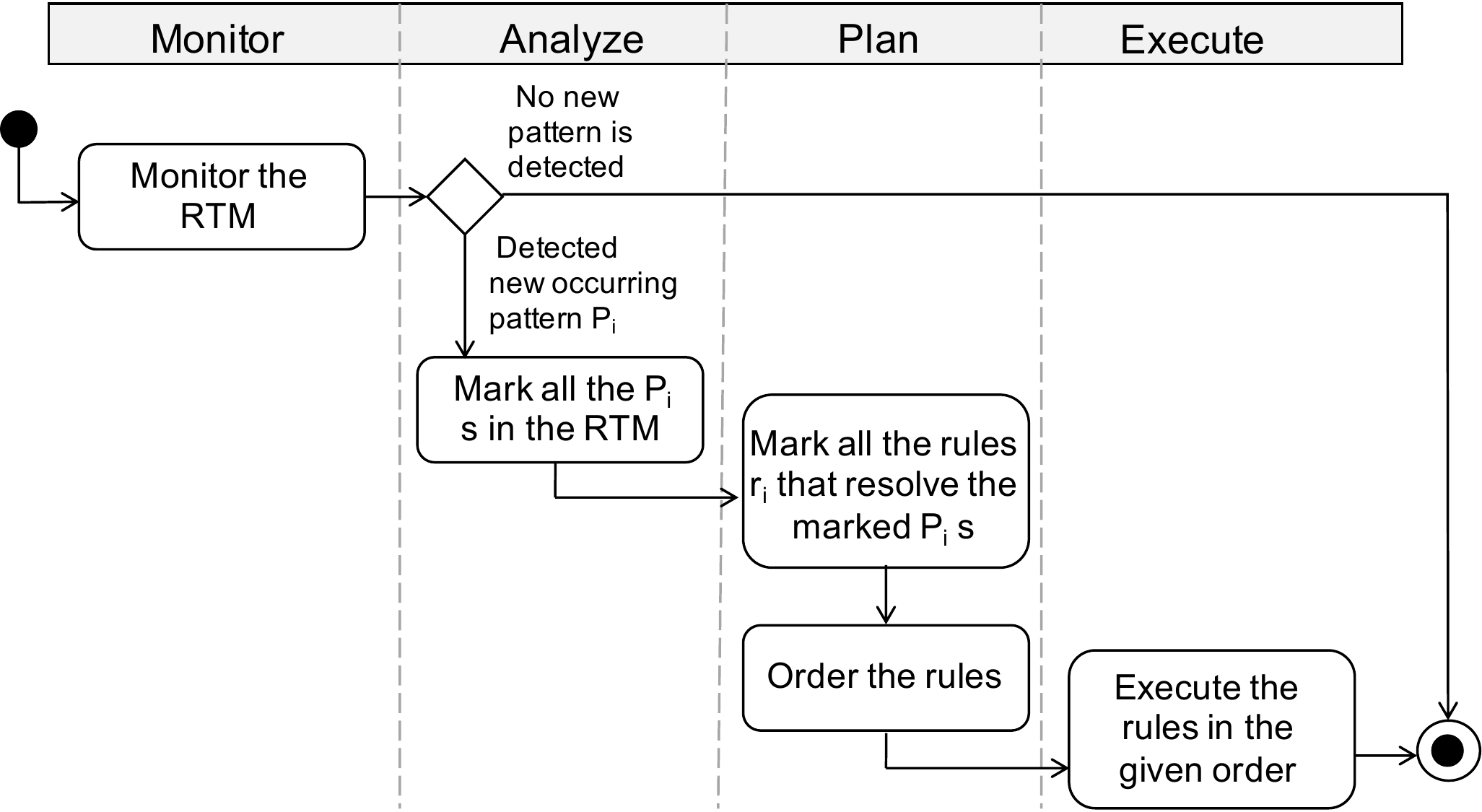}
  \caption{Steps for the Different Phases of the MAPE-K Loop.}
  \label{fig:MAPE}
  \end{centering}
 \vspace{-6mm}
\end{figure}

In contrast, the proposed utility-driven, rule-based scheme determines the impact of each rule application on the utility at runtime. It derives in a greedy manner from these impacts an optimal sequence of rule applications concerning the adaptation decisions. This scheme is realized by a \mbox{MAPE-K} feedback loop as shown in Fig.~\ref{fig:MAPE} and discussed in the following.

\subsection{Monitor}\label{subsec:m}
\noindent
During monitoring, change events emitted by the system are observed and reflected in the runtime model, that is, the model is updated to represent the current system configuration~\cite{Vogel+2009,VogelNHGB10}. In our example, we observe the life cycle \elem{state} of a component (e.g., to monitor whether a component has stopped, crashed, or been removed) and \elem{Failure}s such as exceptions that occur when using a \elem{ProvidedInterface} (cf.~Fig.~\ref{fig:metamodel}).

\subsection{Analyze}\label{subsec:a}
\noindent
In the analysis phase, the observed changes are analyzed to detect negative patterns (issues) in the model. This updates the known set of matches for issues. New matches are determined through applying an event-property-change mechanism and all old matches have to be checked whether they are still valid.

As a first step, we compute the utility incrementally rather than for each configuration anew.
Given a former runtime model $G$ and an updated version $G'$, the set of new occurrences for utility patterns  are $M_i^{new} = M_i(G') \backslash M_i(G)$. Similarly, $M_i^{del} = M_i(G) \backslash M_i(G')$ captures the matches for patterns that are no longer valid. 
We can therefore define the changes of a utility function $U(G)$ accordingly by a utility change function $U_\Delta(G',G)$ as $U(G') - U(G)$ as: 
\begin{equation}\label{equation:incr}
-
  \sum_{i=1}^{n}
      \sum_{m \in\!M^{del}_i}
        \!\!\!\!\!U_i(G,m)
+
  \sum_{i=1}^{n}
      \sum_{m \in\!M^{new}_i}
        \!\!\!\!\!U_i(G',m)
\end{equation}

Besides computing the decrease or increase in utility, we keep track of the identified issues that need to be resolved. 
Considering our focus on self-healing, all the architectural utility patterns that need to be matched and resolved are the negative patterns $\Pa^- = \{ P^-_{k+1}, \dots, P^-_n \}$.

For this purpose, the analyze phase adds \elem{Annotations} to the runtime model. It checks the model for occurrences of negative patterns, which are then annotated as \elem{Issue}s pointing to the \elem{affectedComponent} (Fig.~\ref{fig:metamodel}).
As issues we consider 
crashes (\elem{CF1}) and 
unplanned removals (\elem{CF3}) of components, 
occurrences of \elem{Failure}s (\elem{CF2}), 
and connector crashes (\elem{CF4}).

\begin{figure}[t]
\begin{centering}
  \includegraphics[scale=\scalefactor]{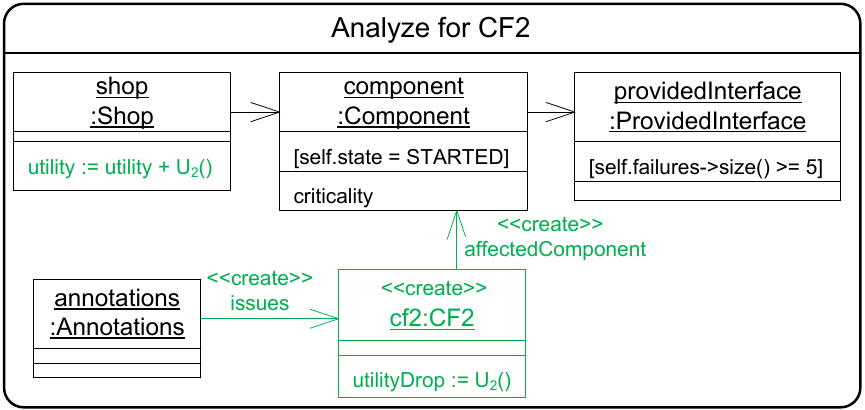}
  \caption{Annotating a Negative Architectural Utility Pattern.}
  \label{fig:A}
  \end{centering}
\vspace{-6mm}
\end{figure}

Fig.~\ref{fig:A} shows an analysis rule realized by a \textit{story pattern}~\cite{FNTZ98_ag} that detects the negative pattern shown in Fig.~\ref{fig:Antipatt}. The occurrence of the negative pattern results in a drop in the utility of the shop by $U_2$.  The story pattern \elem{create}s the \elem{CF2} annotation including the \elem{utilityDrop} pointing to the affected component to be used later on in the feedback loop. 
Here, we omit the details to avoid multiple annotations for the same match (issue). 

\subsection{Plan}\label{subsec:p}
\noindent
Based on the annotations representing new or remaining issues in the form of matches $m$ for negative patterns, our approach incrementally proceeds during the planning phase by
1)~computing the set of all possible rule applications,
2)~selecting for each issue the best rule application based on computations of the impact on utility and costs, and
3)~finally ordering the best rule application for all issues to minimize the lost reward.

\subsubsection{Compute All Possible Adaptation Rule Matches}
The rule-based adaptation is based only on rule applications that lead to an improved utility. 
For this case, we will show that rules must always be linked to the negative patterns and that knowing the matches for these patterns will allow us to compute all relevant adaptation rule matches incrementally.

For any adaptation scheme that is based on the outlined utility function defined by means of patterns where considering our focus on self-healing, all the patterns that need to be matched and resolved are the negative patterns, the following observations must hold:
(1) If there are no occurrences of the negative patterns, then there is no need for adaptation and no improvement of the utility is possible. 
(2) Any possible improvement of the utility must necessarily resolve found occurrences of the negative patterns as otherwise no improvement of the utility would be possible. 

Consequently, we can safely assume that (A1) for any in-place model transformation rule $r_j = (P_j, \dots)$ in the adaptation rule set $\Re$ must hold that a negative pattern $P^-_i$ exists such that any match $m_j$ for $P_j$ includes a match $m_i$ for $P^-_i$. Otherwise, $r_j$ could be enabled even though no utility improvement can be achieved which would contradict observation (1). It can be the case that the initial match condition of the rule $r_j$ might require more context and be more restricted compared to the match condition for the negative pattern, but in the presented example in Fig.~\ref{fig:Antipatt} and~\ref{fig:A}, both conditions are exactly the same. 

Furthermore, we can plausibly assume that (A2) for the rule $r_j = (P_j, \dots)$ in the rule set $\Re$ and any match $m_j$ for $P_j$ and the included match $m_i$ for the related negative pattern $P^-_i$ holds that applying $r_j$ for $m_j$ will make the match $m_i$ for $P^-_i$ invalid. Otherwise, $r_j$ would not handle the found occurrence of the negative patterns $P^-_i$ and thus would not lead to the improvement of the utility we would expect according to observation (2).
To keep our considerations simple, we consider the case where each rule covers exactly one negative pattern. 
Based on these assumptions, we can compute all matches for rules incrementally if for the related negative pattern $P^-_i$ the set of new matches $M_i^{new}$ is given.

\begin{figure}[t]
\begin{minipage}{0.5\columnwidth} 
\center
   \includegraphics[scale=\scalefactor]{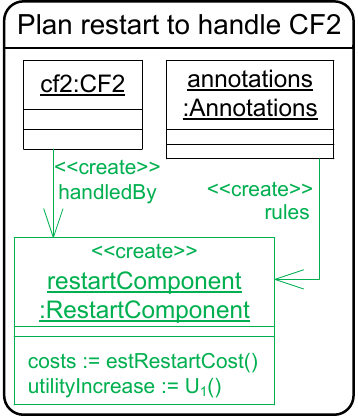}
	\end{minipage}%
	\hfill
	\begin{minipage}{0.5\columnwidth}
	\center
	\includegraphics[scale=\scalefactor]{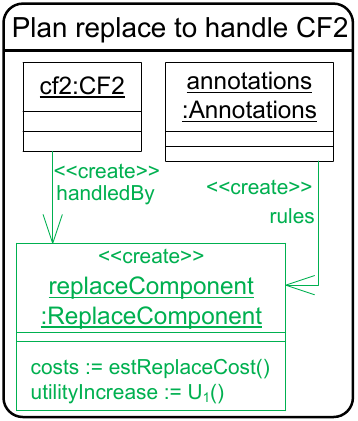}
	\end{minipage}
	\caption{Rules for Planning an Adaptation.}
	\label{fig:P}
\vspace{-5mm}
\end{figure}

Fig.~\ref{fig:P} illustrates a simplified view of a planning rule for the example to repair \elem{CF2}. Following the analysis phase described in Fig.~\ref{fig:A}, the rule set $\Re$ is checked to find all the rules that can extend the occurred negative pattern with adaptation rules such as \elem{RestartComponent} and \elem{ReplaceComponent} that can handle the corresponding issue such as \elem{CF2}. 
The cost estimation functions \elem{estRestartCost()} and \elem{estReplaceCost()} compute the costs of executing each rule to the system under adaptation, for instance, to restart or replace a component in the system.

In general and following the MAPE-K cycle, performing an adaptation consists of a planning and an execution part. The planning part decides which adaptation rule among all possible ones should be applied, while the execution part actually applies the selected rule to prescribe an adaptation in the runtime model that is synchronized to the running system under adaptation (cf. causal connection in Section~\ref{subsec:runtime-model}).

For our example, the planning phase addresses the identified issues by selecting for each of them the adaptation rule to be executed such that it enriches the model with \elem{Rule}s that handle the identified \elem{Issue}s. These rules are finally enacted by the execute phase. As adaptation rules, we consider restarting, redeploying, and replacing components as well as recreating connectors (see Fig.~\ref{fig:metamodel}). For the redeployment, there exists two variants. The light-weight variant keeps the latest configuration of the component to be redeployed, while the heavy-weight variant adapts the configuration of the redeployed component.

In general, a specific issue such as \elem{CF2} can be handled by multiple rules such as restarting or redeploying the affected component. Thus, the planning must decide which rule should be applied. Likewise, if multiple issues occur at the same time, the planning must decide which issue should be handled first.

\subsubsection{Select Adaptation Rule Match for Each Negative Pattern Match Based on the Impact on Utility and Estimates for Costs}

To determine the best adaptation rule for each found occurrence of any negative pattern, our approach determines the impact of adaptation rules on the utility for each match. 

For a single rule $r_j = (P_j,\dots)$ where $P_j$ extends the negative pattern $P^-_i$, it holds that 
each time $r_j$ is applied to $m_j$ then the match $m_i$ for a negative pattern $P^-_i$ is removed (see A2 discussed previously). We further assume that 
(A3) $r_j$ does not result in any new match or removed matches besides $m_i$ for any negative pattern. 
Then, we conclude for any $G,\,G'$ that results from applying rule $r_j$ to $G$ for match $m_j$ ($G \to_{r_j,m_j} G')$:
\begin{equation}\label{eq:delta}
  U^{r_j}_\Delta(G,m_j)
:=
  U_\Delta(G',G) 
=
        U^-_i(G,m_i) 
\end{equation}
Thus, for the discussed form of rules for which (A1) to (A3) hold, we can locally compute their impact on the utility.

If further the assumption (A4) holds that $r_j$ does not affect any utility sub-function for any match $m_k$ for another negative pattern $P^-_k$, 
then applying a rule $r_j$ for a match $m_j$ does not affect the impact on the utility for any other rule $r_k$ and match $m_k$. Thus, if (A1) to (A4) hold, we can independently and locally compute the utility impact of each match of a rule. 

There can be cases that the side effect of applying a rule~$r_j$ (i.e., $G \to_{r_j,m_j} G'$) results in new matches for one or more positive patterns. In such cases, the impact on the utility by the corresponding positive utility sub-function of these matches is added to $U^{r_j}_\Delta(G,m_j)$ in equation\,(\ref{eq:delta}). 
For this purpose, it must hold that all the potential positive patterns are completely within the scope of the application condition and side effect of $r_j$ and do not match only partially (A5). 
Otherwise, matches for the positive patterns cannot be enabled by applying $r_j$.
Thus, the impact can be simply considered in $U^{r_j}_\Delta(G,m_j)$ since the resulting formula for the corresponding increase of the utility can be determined at development-time. 
An example for such a case is replacing a local authentication component with an alternative third-party service. Each of the possible alternative services results in a different positive pattern with different utility sub-functions regarding their offered \elem{reliability}. 

Similarly, the execution time for adaptation rules can be estimated by defining a cost function $Cost^{r_j}(G,m_j)$ for each rule application which may depend on the match and its context in $G$. In our example, we have cost estimation functions for each type of rule such as \elem{estRestartCost()} (see Fig.~\ref{fig:P}).

The planning phase then proceeds by first ranking the applicable rules for each pattern match found according to their \elem{utilityIncrease} $U^{r_j}_\Delta(G,m_j)$ and, if this is the same, according to the \elem{costs} $Cost^{r_j}(G,m_j)$. The \elem{utilityIncrease} of each rule is the computed increase of the overall utility as the result of applying the rule. This attribute value is computed based on the runtime computation of criticality, connectivity, and reliability of the \elem{affectedComponent}. The \elem{costs} for each rule is the computed execution time computed at runtime. 

\subsubsection{Order the Execution for All Selected Adaptation Rule Matches}

The final planning step is determining the order in which the issues should be resolved. For this purpose, all rules annotated in the model are sorted regarding their impact on the overall utility divided by the costs. This guarantees in the execution phase that the maximal utility is reestablished as fast as possible and that the lost reward is minimized.

\subsection{Execute}\label{subsec:e}

\begin{figure}[t]
\begin{centering}
  \includegraphics[scale=\scalefactor]{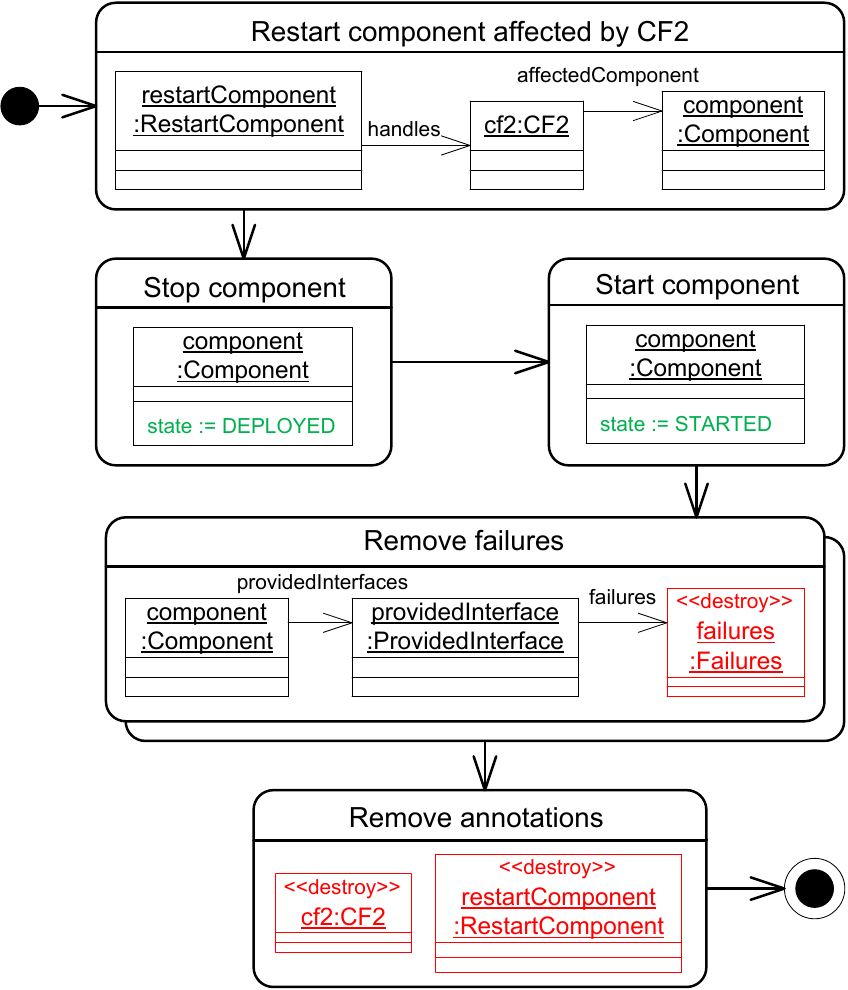}
  \caption{Rule for Executing a Component Restart.}
  \label{fig:E}
  \end{centering}
\vspace{-4mm}
\end{figure}

\noindent
Finally, we combine the beforehand outlined steps such that the utility-driven, rule-based adaptation is achieved. Thus, each detected negative pattern is handled with the most appropriate rule and the rule applications are ordered such that those with the highest impact on utility are executed first.
The execute phase takes over the ordered list of adaptation rule matches from the planning phase and executes them in the given order. 

Fig.~\ref{fig:E} illustrates an adaptation rule that restarts a component to address \elem{CF2}. The control flow within the rule complies with an UML activity diagram (cf. \cite{FNTZ98_ag}). Based on the analysis and planning phases (see Fig.~\ref{fig:A} and~\ref{fig:P}), an adaptation rule, in this case \elem{restartComponent}, has been selected to handle \elem{CF2} affecting the specific \elem{component} (see first node in Fig.~\ref{fig:E}). This component is then restarted (see second and third node). After that, the runtime model is cleaned up by removing (\elem{destroy}ing) the observed exceptions (\elem{failures}) and the annotations for the executed rule (\elem{restartComponent}) and the handled  issue (\elem{CF2}). 

\section{Analysis and Discussion of the Approach}\label{sec:analysis}
\noindent
We now analyze and discuss the computational effort and the optimality of the resulting utility and reward of our approach.

\paragraph{Incremental Computational Effort}
\noindent
If the patterns to be matched as well as the rules do only contain links for associations with fixed small upper bounds smaller than a constant, we can conclude the following: the outline adaptation scheme requires only incremental computational effort for finding new matches for each analysis rule to annotate negative architectural utility pattern, to extend such matches for an adaptation rule, or to check for old issues or rules whether they are still matched.
As we have only a small finite number of rules for analysis and planning, the effort to compute all new issues and determine all related new matches for rules extending these issues for $\Delta$ changes of the architectural runtime model are both in $O(\Delta)$.
The effort to check all old issues and related new matches for rules extending these issues for $\Delta'$ old unprocessed changes of the architectural runtime model are both in $O(\Delta')$.
Thus, the analysis and planning activities only require an incremental computational effort as long as the set of accumulated unprocessed changes do not grow unbounded.

\paragraph{Optimality of the Adaptation}
\noindent
We know that as a result of executing a selected adaptation rule, a maximal increase of the utility is guaranteed. Due to assumption (A2), the rules will remove the match $m$ and, due to assumptions (A3) and (A4), no other matches for issues are affected such that the overall increase after applying all selected rules and matches must be maximal.
Furthermore, the ordering of the rules and determined matches leads to the maximal reward. 
Consequently, for our approach holds that the resulting utility after the complete adaptation is maximal and so is the reward for the chosen sequence of executing the adaptation rules.

\paragraph{Limitations}
\noindent
Rules that are not triggered by any issue or that do not resolve any issue and thus have no impact on the utility do not make any sense so that the assumptions (A1) and (A2) are justified in general. 
Similarly, rules that cause new issues are not helpful and therefore could also be excluded (see first part of (A3)).
However, it will not always be the case that rules do not impact other issues than the ones they should handle (see second part of (A3) and (A4)), for instance, when due to resource limitations the planned sequence of rule applications can not be fully executed such that not all of the issues can be repaired. It can also be the case that rules do not completely cover the positive patterns (see (A5)). 
In these cases to avoid any interferences, the design of the rules could be revised based on an analysis of the rules for conflicts. 
In particular for (A5), this is always possible by splitting the rule into multiple ones taking a larger context into account such that the overlap with the positive patterns is always included. 
If this is not possible or feasible, the \emph{receding horizon} concept from model predictive control \cite{Angelopoulos+2016} of $k$ can be employed where only the first $k$ steps of the plan are realized before the further steps are re-planned anew. By using a receding horizon of $k=1$, we can ensure that the repeated planning steps take the effects of the executed rules into account.

\section{Experimental Evaluation}\label{sec:evaluation}
\noindent
To evaluate our approach, we use a simulator of \mRUBiS~\cite{mRUBiS}, a variant of the common RUBiS that is frequently used for validating self-adaptation mechanisms~\cite{Patikirikorala+2012}. Having fault injection capabilities, the simulator emulates the failures in the system by reflecting them in the runtime model as it would be otherwise done by monitoring the faulty system. \mRUBiS hosts different numbers of shops ($1$\,to\,$1000$), each containing $18$ components with a different \elem{criticality} and \elem{connectivity}.\footnote{The experiments and simulations have been conducted on a machine with OS\,X $10.10$, Intel processor $ 2.6\,GHz$ core $i5$, and $8\,GB$ of memory.} The overall utility of a shop is the sum of the sub-utilities of all the components in the shop. As described in Section~\ref{sec:arch-based}, we equipped \mRUBiS with a MAPE-K feedback loop. The three issues \elem{CF1}, \elem{CF2}, and \elem{CF3} are the negative patterns that affect the system. The rule set $\Re$ includes the adaptation rules each representing a repair plan. Each rule has two attributes, \elem{costs} and \elem{utilityIncrease} (see Fig.~\ref{fig:metamodel}). \elem{Costs} refers to the execution time of the rule and \elem{utilityIncrease} is the impact on the utility of the \elem{affectedComponent} when applying the rule. 

We validate the optimality of our scheme with analytical experiments (Section~\ref{subsec:Aresult}). 
Moreover, we investigate the scalability and performance of our approach in a comparative study with two alternative approaches (Section~\ref{subsec:Eresult}).  
Thus, the study compares three solutions:

  \subsubsection{Static Approach}\label{subsubsec:static}
  This approach is purely rule-based and uses static priorities without any utility function.
  Thus, the \elem{costs} and \elem{utilityIncrease} of the rules are defined at design time so that for each \elem{CF} the repair rule is selected staticly. The \elem{utilityDrop} caused by each \elem{CF} is also estimated at design time which leads to a fixed order in which the issues are resolved. 
  \subsubsection{Solver-based Approach}\label{subsubsec:solver}
  This approach is purely utility-based and uses the IBM ILOG CPLEX constraint solver~\cite{Cplex} for planning. 
  Specifically, it uses the utility function described in equation\,(\ref{eq:utility}) for the sequence of rule applications as its objective function. The tasks of assigning proper adaptation rules to each \elem{CF} and ordering them are defined as optimization problems. This approach maximizes the objective function as the overall utility of the system after each decision. 
  \subsubsection{Utility-driven Approach}\label{subsubsec:dynamic}
  
 Our approach computes the impact of different adaptation rules at runtime using the utility function shown in equation\,(\ref{eq:utility}) and selects the one with largest impact on the overall utility. The order in which \elem{CF}s are addressed and the proper adaptation rule to resolve the \elem{CF}s are decided based on the runtime observations regarding the \elem{affectedComponent} and the utility drop caused by the \elem{CF}s. This approach is referred to as \textit{u-driven} in the following. 

Thus, the three approaches have different planning phases while they share the same incremental behavior---as suggested for our approach---for the other phases of MAPE-K.

\subsection{Analytical Experiments}\label{subsec:Aresult}
\noindent
The conducted experiments for analytical purposes are set to separately evaluate the two main steps of our approach. In these experiments, we consider \mRUBiS with 100 shops (1800 components). The experiment starts with occurrences of three failures of type \elem{CF1}, \elem{CF2}, and \elem{CF3} causing the utility of the system to drop. Utility drops are followed by three MAPE executions. During each MAPE cycle, we consider a receding horizon of size one resolving one \elem{CF} in one MAPE execution. As the effort of the repeated planning step is negligible due to its incremental nature, a receding horizon of $1$ will take the effects of the executed rules into account, even though assumptions (A3) to (A5) do not hold.

Here, we investigate that the u-driven approach makes the optimal decision during incremental rule matching for \elem{CF}s by selecting the rule that results in the maximum increase in the overall utility. In contrast, the static approach fails to do so and hence is non-optimal. We also show how the order in which the adaptation rules are executed impacts the achieved reward. 

When a match for an issue is detected, our approach computes the \elem{utilityIncrease} and \elem{costs} of all possible matches among the adaptation rules. The effect of the increase in the utility achieved by applying each rule remains in the system as long as the same component is not affected again by another issue. However, the costs of applying a rule has only a short time effect on the overall utility and defines when the expected increase of the utility can be realized. Thus, in our approach, rules with the highest \elem{utilityIncrease} are prior to those with lower increase but less \elem{costs}. The type of the occurred issue and the specific component that is affected by the issue determine the \elem{utilityIncrease} and \elem{costs} of the rules.

 \begin{figure}[t]
\begin{centering}
  \includegraphics[width=0.90\linewidth]{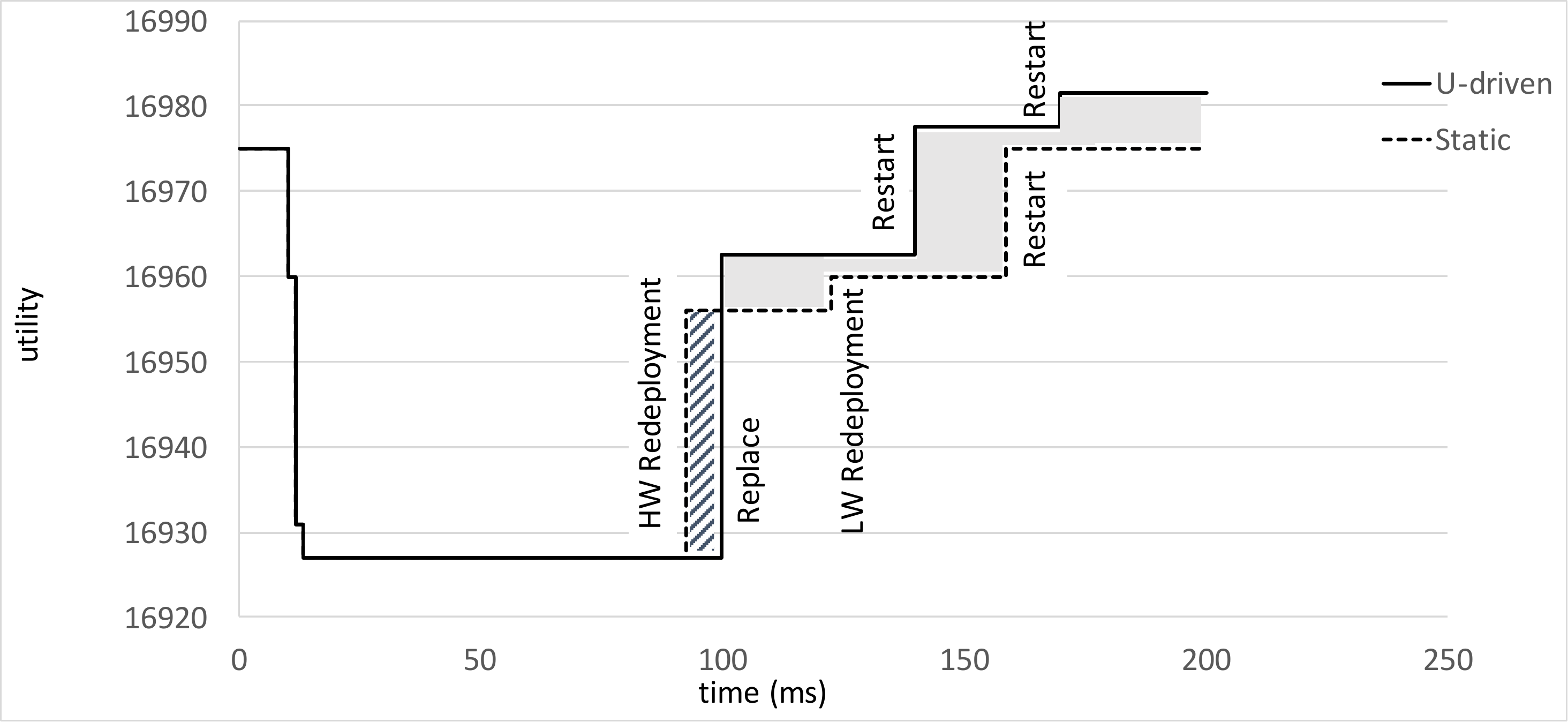}
  \caption{Lost Reward Due to Non-optimal Decisions.}
  \label{fig:stu}
  \end{centering}
\vspace{-7mm}
\end{figure}

Fig.~\ref{fig:stu} describes a case where the static approach fails to reach the maximum utility due to non-optimal rule selection. During the first MAPE cycle, both approaches select \elem{CF3} to be resolved first. The static approach performs a \elem{Heavy Weight (HW) Redeployment} while the u-driven approach \elem{Replace}s the affected component and reaches a higher utility. 
Here, the static approach selects a rule with less \elem{costs} and manages to have the utility increase faster than the u-driven approach but obtaining a considerably smaller reward (the hachure region after the first increase). The impact of this non-optimal rule selection remains in the system during the whole experiment and results in a lower reward for the static approach equal to the area of gray colored regions.
In the second MAPE cycle, the static approach resolves \elem{CF1} by a \elem{Light Weight (LW) Redeployment} while the u-driven approach decides to resolve \elem{CF2} by a \elem{Restart} of the component which has higher impact on the overall utility. 
In the third MAPE cycle, the static approach resolves \elem{CF2} by a \elem{Restart} and reaches the same increase in utility as the u-driven approach in the second MAPE cycle, but with a delay and thus with a lower reward as the utility is lower during the whole time. 
The u-driven scheme repairs \elem{CF1} by a \elem{Restart} in the last MAPE cycle. 
The static approach is slightly faster than the u-driven approach due to avoiding all the runtime computations. The gray and hachure regions represent respectively the lost and gained utility of the static compared to the u-driven approach. The additional utility gained by the static approach due to less overhead and choosing the cheaper \elem{HW Redeployment} over the \elem{Replace} rule does not compensate for the loss of reward due to making non-optimal decisions.

\begin{figure}[t]
\begin{centering}
  \includegraphics[width=0.99\linewidth]{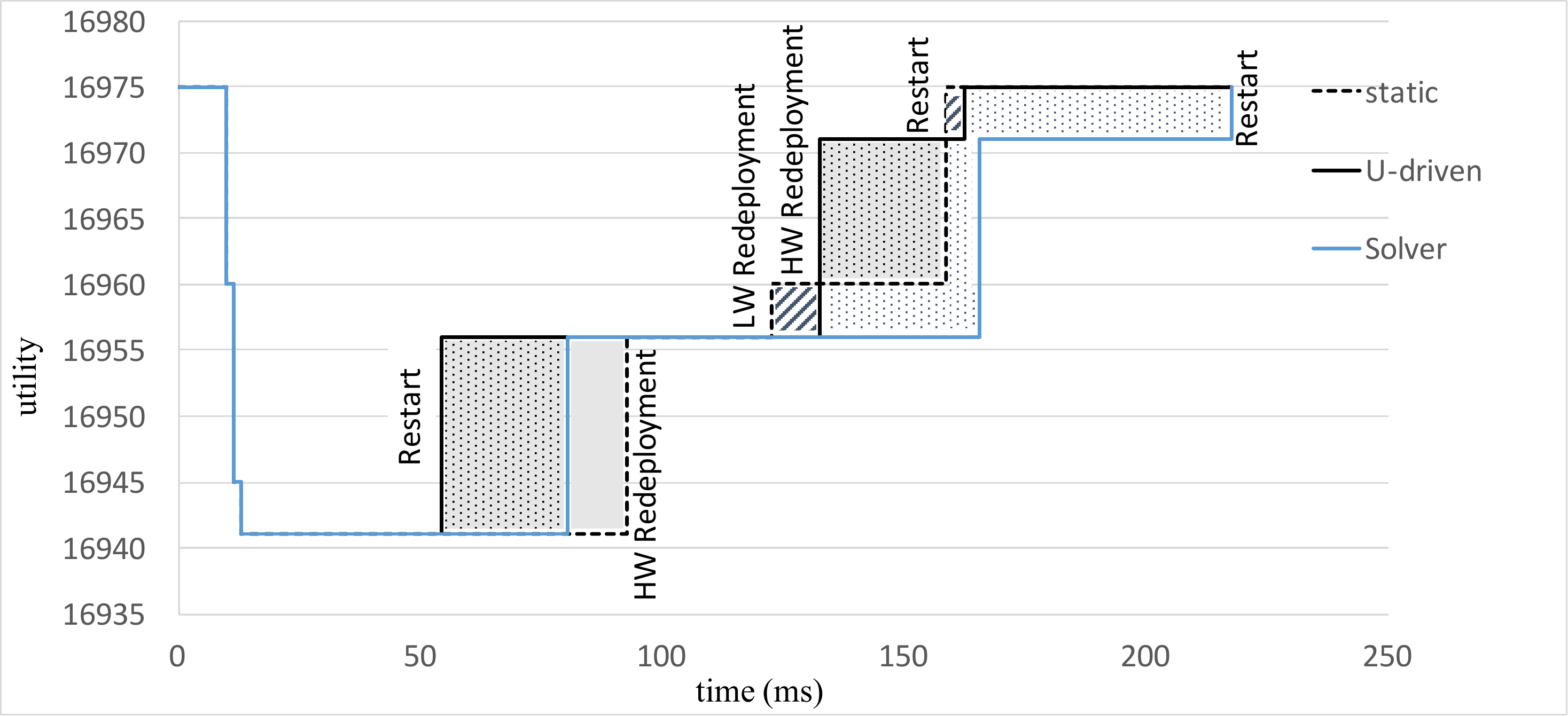}
  \caption{Lost Reward Due to Wrong Ordering and Overhead.}
  \label{fig:sttime}
  \end{centering}
  \vspace{-8mm}
\end{figure}

To back our claim for optimality of the u-driven scheme, our approach executes the adaptation rules in the optimal order such that the maximum utility over time is achieved. We investigate this issue in Fig.~\ref{fig:sttime}. The order in which our scheme resolves the issues is such that those resulting in a higher increase in utility are prior to those with lower impacts. The static approach decides for the order at design time. This can be done considering the type of the potential issues. A reasonable order of the three issues in our example is: 
severe crashes (i.e., unplanned removals) of components (\elem{CF3}),
crashes of components that, however, might still be operating to a certain extent (\elem{CF1}), and
occurrences of \elem{Failure}s such as exceptions (\elem{CF2}).
This ordering fails to take into account the utility of the \elem{affectedComponent} which is a function of criticality, connectivity, and reliability. 
Such properties can dynamically change such that they are only known at runtime and cannot be foreseen at design time. Fig.~\ref{fig:sttime} illustrates a case where the static approach fails to address the issues in the right order. Despite the fact that both, the u-driven and static approaches achieve the same final utility, which is not necessarily always the case (cf.~Fig.~\ref{fig:stu}), the static approach loses reward equal to the gray regions and gains only a slight improvement due to the lower overhead in planning time (hachure region).

Considering Fig.~\ref{fig:sttime}, both, the u-driven and static approaches repair \elem{CF3} in the first MAPE cycle.
The static approach applies a \elem{HW Redeployment} while the u-driven approach \elem{Restart}s the affected component reaching the same utility but with considerably less cost. In this MAPE cycle, the static approach selects a rule with a similar \elem{utilityIncrease} to the one selected by the u-driven approach but with higher \elem{costs} such that it loses utility equal to the area of the first gray region. 
In the second MAPE cycle, the static approach resolves \elem{CF1} by a \elem{LW Redeployment} while the u-driven approach decides to resolve \elem{CF2} by a \elem{HW Redeployment} since solving \elem{CF2} has a higher impact on the overall utility. 
In the third MAPE cycle, the static approach resolves \elem{CF2} by a \elem{Restart} and reaches the same increase in utility as the u-driven approach, but loses utility over time due to the wrong execution order. The u-driven scheme saves the repair of \elem{CF1} by a \elem{Restart} for the last MAPE cycle since in this case \elem{CF1} has less impact than \elem{CF2} and \elem{CF3} on the utility. 

We conducted the same experiment to compare the solver-based and u-driven approaches.
As depicted in Fig.~\ref{fig:sttime}, both approaches make similar decisions regarding both rule matching and ordering of the adaptation rules and reach the optimal configuration. However, the solver-based approach reaches it after a considerable delay due to its computational overhead for planning, which depends on the size of the architecture and number of the issues. 
It is visible that despite the fact that both approaches select the same rules each time with similar \elem{costs} and have the same ordering of the issues, the solver approach finishes with a large delay compared to the u-driven approach.


\subsection{Experiments for Performance}\label{subsec:Eresult}

\begin{table}[b]
\vspace{-4mm}
\begin{centering}
	\caption{Planning Time of the Approaches in (ms).}
  \includegraphics[width=0.99\linewidth]{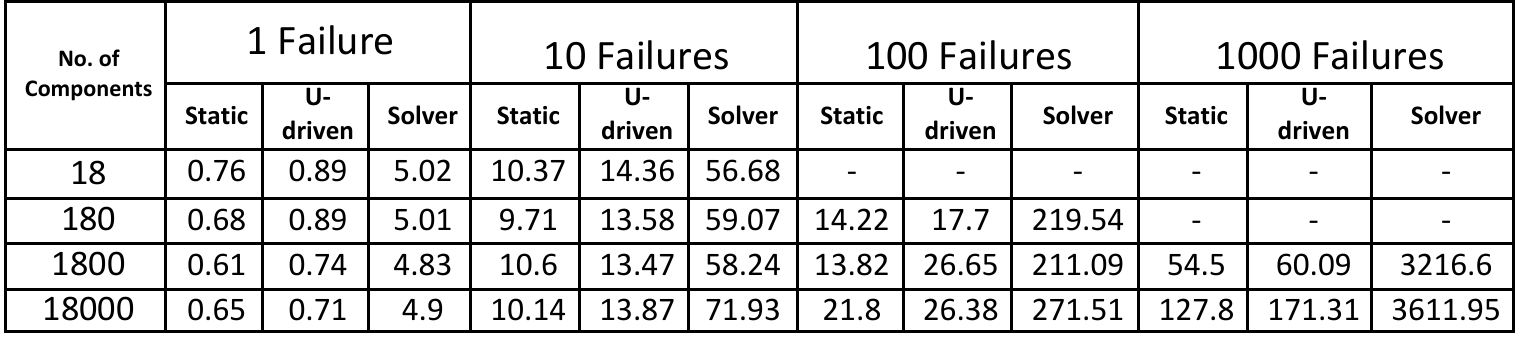}
  \label{fig:Table}
  \end{centering}
\end{table}

\noindent
To compare the performance of the approaches, we tested them on \mRUBiS with $4$ different sizes of the architecture and with scenarios in which $1$, $10$, $100$ and $1000$ failures occur. In each scenario, the equal number of failures of all \elem{CF} types are injected. The measurements are repeated at least $300$ times and until the standard deviation is $5\%$ or lower. The measurements were conducted following benchmarking guidelines~\cite{d62f204fc9d94663a04836dc217ded08}. All three approaches are tested with the same scenarios. Since the decision-making process takes place during the planning phase and the potential overheads due to runtime computations are reflected there, we only present the data for the planning phases of the approaches in Table~\ref{fig:Table}.

For the measurements, we only consider meaningful combination of architecture size and number of failures. Thus, we do not inject a large number failures to small architectures. Therefore, we do not present any data where more than $10$ ($100$) failures occur in a system with $18$ ($180$) components. 

As the number of failures in the system increases, the planning time of all approaches increases as well. However, this growth is more drastic for the solver as the size of the optimization problem is growing. Based on the presented data, the solver-based approach requires between $295\%$ to $5953\%$ more time for planning than the u-driven approach [42 to 3156 (ms)]. The solver approach always reaches the same optimal configuration as our u-driven scheme, but it can have an extreme overhead in the case of a large number of issues.
The additional planning time required by the u-driven approach compared to the static approach varies between $9\%$ to $92\%$, [0.06 to 12.8 (ms)]. 

\section{Related Work}\label{sec:related}
\noindent
As the related work of this work, we investigate how the trade-off between long-term planning and aiming for an optimal repair or settling for a quick and efficient adaptation has been practiced in the field of self-adaptive software.

On one end of the spectrum, there are optimization-based approaches employing runtime reasoning. An objective function is computed at runtime to investigate all the potential decisions, thus encountering scalability and efficiency issues \cite{1691383, 5069076}. Employing utility functions and utility-driven decision-making schemes have been extensively investigated \cite{Kramer&Magee2007,Franco+2016}.
MUSIC \cite{RouvoyEtAl09} is a planning-based middleware for component-based application that plans the adaptation by exploiting the characteristics of component implementations for the software architecture.
\cite{5069076} applies a reinforcement learning-based approach for on-line planning. \cite{Esfahani+2013} solves an optimization problem to find the optimal set of features that maximizes the utility via a learning-based method.
Solving an optimization algorithm for each reconfiguration at runtime causes large overheads.
The outlined utility-driven approaches pursue a search-based optimization in the solution space that often do not scale well for complex systems. Such approaches manage to find the optimal configuration but there is no guarantee to reach the result within a reasonable time, for instance, when quickly needing a repair plan. In \cite{TG04_ag}, we suggested to reduce the search space to speed up repair and avoid too long delays. 
However, the approach proposed here computes the utility for each potential adaptation strategy in an incremental scheme taking into account the current issues that affect the achieved utility. Therefore, the approach is scalable and does not have to restrict the search space for the considered self-healing setting. 

On the other end of the spectrum, there are pure rule-based approaches~\cite{1537890}. They are recognized to be efficient and stable in predictable domains and support the early validation~\cite{1691383}. These approaches provide a quick recovery from a goal violation, however, they often result in sub-optimal solutions since they ignore the scenarios that are unforeseen at design time~\cite{OwenCheng2008}. Rainbow applies utility theory in combination with a stochastic model of the possible outcomes of the reasoning process~\cite{2012ChengStitch}.
While in our approach the utility of the adaptation rules is dynamically computed at runtime, Rainbow considers the success rate of adaptation rules in the past to rank them.

The proposed approach is distinguished from the existing work as it is fast and optimal since it employs adaptation rules and does not struggle with scalability issues. Employing a utility function guarantees optimal adaptation decisions on top of the applied rule-based scheme. However, unlike the optimization-based approaches, the incremental manner of computing the utility function over the patterns makes the approach scalable for large complex systems.

\section{Conclusion and Future Work}\label{sec:conclusion}
\noindent
Achieving optimal adaptation decisions online within a reasonable time is an important challenge addressed by this work. We presented a novel approach to improve the self-healing reward by combining utility-driven and rule-based adaptation at the architectural level to achieve the benefits of each of them. The approach addresses the requirements of scalability and optimality regarding the utility computation. 
Our experiments demonstrate that our approach results in significantly improved reward compared to an alternative static approach while only having a negligible overhead. 
The comparison of our approach to a solver-based solution shows that both perform optimal adaptation decisions while our approach drastically reduces the computation efforts for planning self-adaptation.

However, the presented approach has some limitations that we plan to address in future work. 
This includes weakening the assumptions, supporting more complex utility functions, and studying how to support other capabilities of self-adaptive software such as self-configuration and self-optimization.
 
\bibliographystyle{abbrv}
\bibliography{references}

\end{document}